\documentclass[aps,prd,preprint,groupedaddress,showpacs,amsmath,amssymb]{revtex4}
\usepackage{graphicx}
\begin{document}
\title{The Duality of Time Dilation and Relative Velocity}
\author{Hunter Monroe}
\date{\today}
\thanks{
In honor of the retirement from Davidson College of Dr. L. Richardson King, an extraordinary teacher and mathematician. Email: relpap at huntermonroe.com. This benefits from comments on an earlier version by V. Frolov, B. Iliev, L.R. King, A. de la Macorra, A. Shatskiy, S. Yazadjiev, and participants in a seminar at Davidson College. The views expressed in this paper are those of the author and do not necessarily represent those of the International Monetary Fund or IMF policy.}
\affiliation{International Monetary Fund,Washington, DC 20431}
\newcommand{\dd}{d}
\newcommand{\ff}{{\rm ff}}
\newcommand{\vff}{\beta}
\hyphenpenalty=3000
\begin{abstract}
Time dilation $\frac{1}{\sqrt{1-v^2}}$ and relative velocity $v$ are observationally indistinguishable in the special theory of relativity, a duality that carries over into the general theory under Fermi coordinates along a curve (in coordinate-independent language, in the tangent Minkowski space along the curve). For example, on a clock stationary at radius $r$, a distant observer sees time dilation of $\frac{1}{\sqrt{1-v^2}}=\frac{1}{\sqrt{1-2M/r}}$ under the Schwarzschild metric and sees the clock receding with a relative velocity of $v=\sqrt{2M/r}$ under the Painlev{\'e}-Gullstrand free fall metric. Duality implies that during gravitational collapse, the intensifying time dilation observed at the star's center from a fixed radius $r>0$ is indistinguishable (along a curve) from an increasing relative velocity at which the center recedes as seen from any direction, implying a local inflation.
\end{abstract}
\maketitle
\section{Introduction\label{introduction}}
This note points out that the observational indistinguishability of time dilation $\frac{1}{\sqrt{1-v^2}}$ and relative velocity $v$ in the special theory of relativity is a duality which carries over from the special to the general theory of relativity under Fermi coordinates (along a curve).\footnote{For more information on Fermi coordinates see Nesterov \cite{Nesterov:1999ix} and Iliev \cite{Iliev:1992um}.}

\section{The outcome of gravitational collapse\label{collapse}}



Note that a distant observer sees time dilation at radius $r$ of $\frac{1}{\sqrt{1-v^2}}=\frac{1}{\sqrt{1-2M/r}}$ which is equivalent to velocity of $v=\sqrt{2M/r}$ under Fermi coordinates for an observer free falling from infinity, that is, under the Painlev{\'e}-Gullstrand metric (see for instance Taylor and Wheeler \cite{Taylorandwheeler})
\begin{equation}
  d \tau^2 = \left(1 - \frac{2M}{r} \right ) d t^2 - 2 \sqrt{2M/r} dt dr- dr^2-r^2(d \theta^2+ \sin ^2\theta  d \phi^2)
\end{equation}
Hamilton and Lisle \cite{Hamilton:2004au} interpret this metric as a ``river model" of black holes: space itself appears to flow like a river through a flat background, while objects move through the river according to the rules of special relativity. The river flows inward at the Newtonian escape velocity $\sqrt{2M/r}$ reaching the speed of light at the horizon.

The observational indistinguishability of relative velocity and time dilation implies the following interpretation of intensifying time dilation during gravitational collapse. The unbounded intensification of time dilation observed from any fixed radius $r>0$ as an event horizon emerges at the center $r=0$ of a collapsing star suggests that the center is receding at a velocity that increases toward the speed of light.

This intuition can be formalized as follows. Define the following procedure similar to the Global Positioning System (GPS): In the special theory, suppose two clocks are synchronized and initially co-located, and clock $A$ remains at the origin while clock $B$ recedes in the $x$ direction at velocity $v$. An observer at clock $A$ can determine the exact location of clock $B$ in his coordinates by "GPS" simply by reading that clock: the observer at clock $A$ sees time dilation of $\frac{1}{\sqrt{1-v^2}}$ on clock $B$ which implies a velocity $v$, and multiplying $v$ the time passed on clock $A$ gives the $x$ coordinate of clock $B$. GPS can also be applied under Fermi coordinates along a curve in the general theory.

Suppose a distant observer of gravitational collapse $D$ naively applies this procedure to calculate the trajectories of a series of clocks at constant $r$ between $r=0$ and $r= 2M$. The observer $D$ will calculate the trajectories seen in Figure 1 which has the $r$ axis at an angle for an incoming null line, corresponding to $D$'s observations of the clocks. The clocks initially maintain a constant distance at a very early stage of collapse when time dilation is negligible, and the trajectory of a given clock will approach a null line as the observed time dilation on that increases without bound during collapse. Similarly, an observer stationary at $r=2M$ will calculate the trajectories seen in Figure 2. As shown in Figure 2, the relative velocities seen by an observer at $r=2M$ of the clocks located between $r=0$ and $r=2M$ range between 0 and the speed of light.\footnote{As an example of this type of calculation, an observer stationary at $r_2=8M$ ($v=\sqrt{2M/r_2}=1/2$) calculates the relative velocity of a clock at  $r_1=\frac{25M}{8}$ ($u=\sqrt{2M/r_1}=4/5$) using Einstein's velocity composition law as half the speed of light (if $w=1/2$, then $u=\frac{v+w}{1+vw}=\frac{1/2+1/2}{1+1/4}=4/5$).} Because Fermi coordinates exist along the incoming null line to the observer at $r=2M$, this calculation of velocity is accurate along any given null line. The range of velocities 0 and the speed of light implies a bubble-like local inflation of the star's interior even as the star continues to collapse as seen from the outside, which creates room for gravitational collapse to continue. Shatskiy \cite{Shatskiy:2004tq} conjectured that gravitational collapse would lead to such an effect.

The apparent local inflation can be seen in the context of the inflationary universe literature initiated by Guth \cite{Guth:1980zm}, in this case with the false vacuum created by Pauli exclusion effect, and the passage of the event horizon from the center of the star to its surface acting as a phase change. By contrast with the literature on the limiting curvature hypothesis (see the Easson \cite{Easson:2002tg} and references), here a dynamic mechanism which is deduced rather than assumed regulates curvature by lowering it whenever time dilation is significant. Under the limiting curvature hypothesis, Frolov {\it et al} \cite{Frolov:1988vj}\cite{Frolov:1989pf} observe that black holes generate closed universes. Easson and Brandenberger \cite{Easson:2001qf} note that the generation of universes from black hole interiors would solve the horizon, flatness, and structure formation problems and also resolve the information loss and other paradoxes.

Local inflation would appear to avoid the creation of a trapped surface and a singularity as the inevitable outcome of gravitational collapse (see Oppenheimer and Snyder \cite{Oppenheimer:1939ue} and Penrose \cite{Penrose:1964wq}).\footnote{Mitra \cite{Mitra:1999yr} and Leiter and Robinson \cite{Leiter2} also conclude that trapped surfaces cannot occur.} Local inflation would also avoid an inevitable singularity before the big bang, as predicted by Hawking and Penrose \cite{Hawking:1969sw}, because there is no time-reversed trapped surface inside the horizon: past-directed geodesics do not converge but cross the event horizon at distinct points in spacetime.

\bibliography{equivalence}
\begin{figure}
\includegraphics{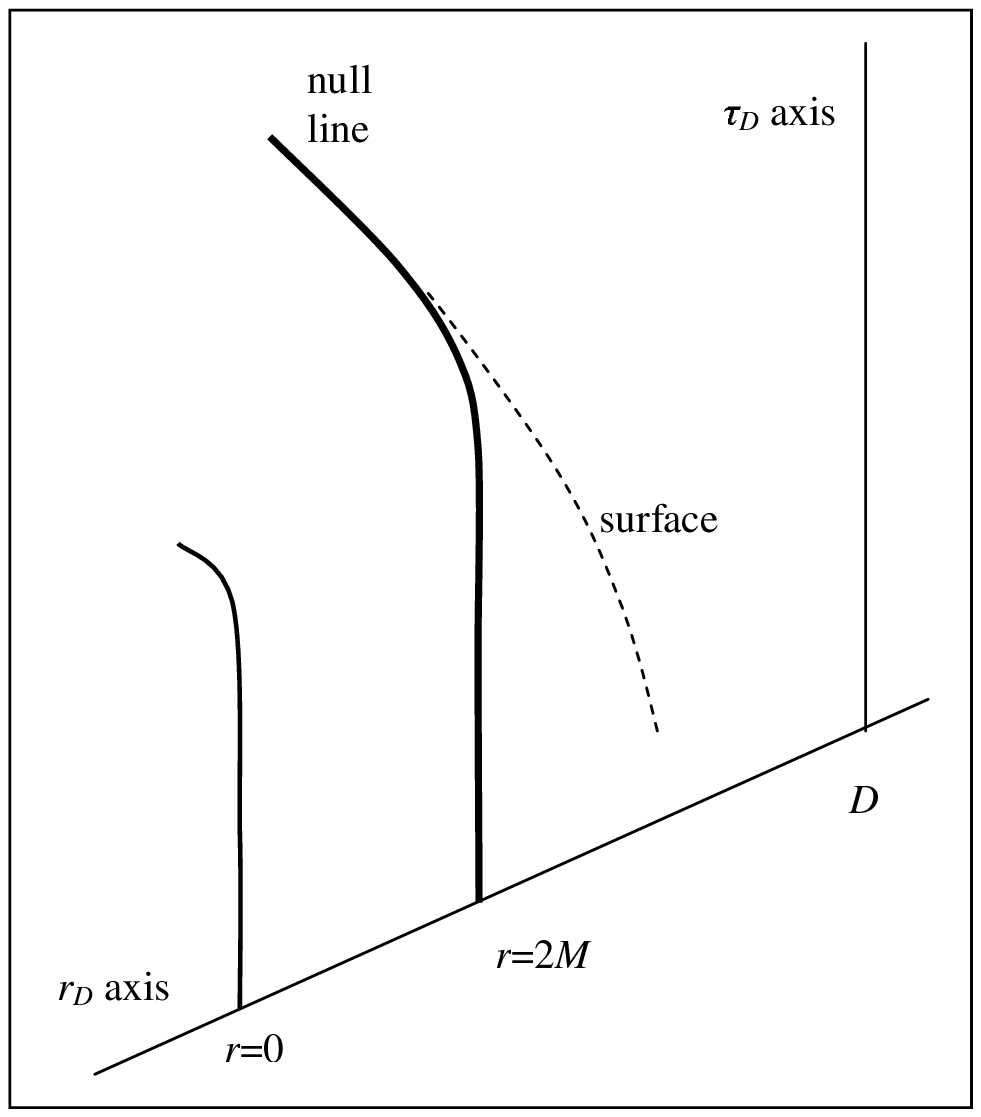}
\caption{\label{fig:epsart1}Trajectories at fixed $r$ calculated by GPS by a distant observer $D$}
\end{figure}
\begin{figure}
\includegraphics{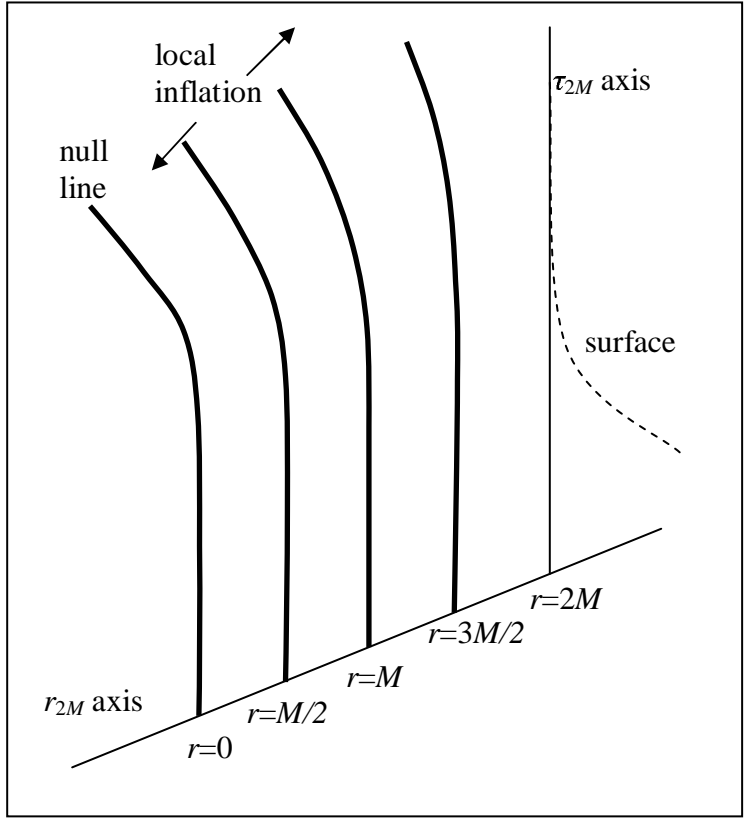}
\caption{\label{fig:epsart2}Trajectories at fixed $r$ calculated by GPS by an observer at $r=2M$}
\end{figure}
\end{document}